\documentclass[prd,aps,showpacs,showkeys,nofootinbib]{revtex4}
\usepackage{amssymb,amsmath,epsfig}
\usepackage{lineno}
\usepackage[usenames,dvipsnames]{color}
\usepackage[bookmarks]{hyperref}
\usepackage[svgnames]{xcolor}
\usepackage{subfigure}
\usepackage{times}
\usepackage{yhmath}
\usepackage{soul}
\usepackage{cancel}

\newcommand{\p}{{\rm p}}

\newcommand{\kb}{k_B}
\newcommand{\vt}{v_{th}}

\begin{document}
\title{Existence of the Chapman-Enskog solution and its relation with first-order dissipative fluid theories}
\author{A. L. García-Perciante$^{1}$, A. R. Méndez$^{1}$, and O. Sarbach$^{1,2}$}
\affiliation{$^{1}$Departamento de Matemáticas Aplicadas y Sistemas, Universidad
Autónoma Metropolitana-Cuajimalpa (05348) Cuajimalpa de Morelos, Ciudad
de México México.}
\affiliation{$^{2}$Instituto de Física y Matemáticas, Universidad Michoacana de
San Nicolás de Hidalgo, Edificio C-3, Ciudad Universitaria, 58040
Morelia, Michoacán, México.}
\begin{abstract}
The conditions for the existence of the Chapman-Enskog first-order solution to the Boltzmann equation for a dilute gas are
examined from two points of view. The traditional procedure is contrasted
with a somehow more formal approach based on the properties of the linearized collision operator.
It is shown that both methods lead
to the same integral equation in the non-relativistic scenario. Meanwhile,
for relativistic systems, the source term in the integral equation adopts two different forms. However, as we explain, this does not lead to an inconsistency.
In fact, the constitutive equations that are obtained from both methods  are shown to be equivalent within relativistic first-order theories. The importance of stating invariant definitions for the transport coefficients in this context is emphasized. 
\end{abstract}
\date{\today}

\maketitle


\section{Introduction and preliminaries}

\label{Sec:Introduction} 
The study of perturbative solutions to the Boltzmann equation in
the dilute gas scenario dates back to work by Hilbert,
specifically in relation to the so called 6th problem \cite{Hilbert,Hilbert6thBobylev}.
In particular, the Hilbert expansion, suitably modified by Chapman
and Enskog, led to one of the most successful methods of solution
for the Boltzmann equation in dilute systems: the Chapman-Enskog (CE)
approximation \cite{Ch-E,cercigniani1}. Despite criticisms on its
mathematical weaknesses, mostly on regimes beyond Navier-Stokes (see
for example \cite{Hilbert6thBobylev,cercigniani1,LaurePaper}), this
method has been shown to be a powerful tool capable of fundamenting
the hydrodynamic equations and delivering constitutive relations for
dissipative fluxes which agree with experimental results in the linear regime.

In the relativistic scenario, the CE approximation was
applied by pioneers in the field such as W. Israel~\cite{Israel63} and S. R. de Groot~\cite{deGrootLibro}. However,
the fact that its traditional form predicts first-order couplings between dissipative fluxes and spatial gradients of the state variables leads to its partial dismissal, based on the findings that such relations give rise to pathological theories for which the equilibrium configurations are generically unstable~\cite{Hiscock}. 
Recently, a new family of first-order theories has been proposed and shown to contain promising candidates to describe high temperature
gases in curved spacetimes. These new theories allow one to consider
a general frame and representation in which constitutive equations feature all possible couplings, including relations which involve time derivatives of the state variables. Under suitable restrictions, these theories have been shown to lead to physically
sound equations~\cite{BDNK_PRX,pK19}. However, despite recent attempts~\cite{bdnk19,rocha22}, their microscopical foundations are still not well understood from the point of view of the CE expansion. For recent progress based on the Hilbert expansion, see Ref.~\cite{rHpK22}. The purpose of this work is to shed new light on this problem. 

As a starting point we consider the Boltzmann
equation, which is an integrodifferential relation describing the
balance between the time evolution of the one-particle distribution function $f$ and the cumulative effects of collisions,
and which we write here in a general form as
\begin{equation}
L_{F}\left[f\right]=Q\left(ff\,'\right).
\label{eq1-1}
\end{equation}
Here $L_{F}$ represents the Liouville operator, giving the total
time derivative of the distribution function in the presence of an
external electromagnetic force (indicated symbolically by $F$), and
$Q\left(ff\,'\right)$ is an integral operator accounting for the
balance of particles in a cell of phase space due to binary interactions, see Refs.~\cite{Ch-E,CercignaniKremer-Book} for details.
The CE expansion considers small perturbations of a local equilibrium state $f^{\left(0\right)}$ such that 
$f=f^{\left(0\right)}\left( 1 + \phi \right)$. Up to first order, $f^{\left(0\right)}$ and $\phi$ satisfy
\begin{equation}
Q\left(f^{\left(0\right)}f^{\left(0\right)'}\right)=0,
\label{2}
\end{equation}
and
\begin{equation}
C\left(\phi\right)=L_{F}\left[\ln f^{\left(0\right)}\right],
\label{2-2}
\end{equation}
where $C\left(\phi\right)$ is the linearization of the operator $Q\left(ff\,'\right)$ at $f^{\left(0\right)}$ (see below for more details). Notice that we assume in this article that the scale of the electromagnetic field is macroscopic. In this work,
we focus on the source term $L_{F}[\ln f^{\left(0\right)}]$ which needs to lie in the image of the operator $C$. We achieve this using two different methods. 

Firstly, the traditional procedure is implemented, which performs an expansion of the convective time derivative term in
the right-hand side of Eq.~(\ref{2-2}) \cite{Ch-E,deGrootLibro,CercignaniKremer-Book}. The coefficients of this expansion are adjusted by integrating the first-order equation, multiplied by the collision invariants, over the momentum space.
This procedure is shown to be equivalent to substituting the time
derivatives of the state variables using the local equilibrium Euler
equations, and renders the source term depending solely on the spatial gradients.

The second, more formal, method follows the lines of Ref.~\cite{LaurePaper} (see also~\cite{LaureLectureNotes}) by projecting the source term $L_{F}\left[\ln f^{\left(0\right)}\right]$
onto the orthogonal complement of the kernel of $C$, which consists of the collision invariants. Since the operator $C$ is self-adjoint when defined in an appropriate Hilbert space, this assures that the source term lies in the image of $C$.

As is shown in the first part of this work, in the non-relativistic case, both methods lead to the same equation and thus to the traditional results in non-equilibrium thermodynamics, including the Fourier and Navier-Newton (or Navier-Stokes) constitutive relations. On the other hand, in the relativistic scenario, the two procedures lead to different versions of the source term. In particular, the second method yields a source term that depends on spatial gradients as well as proper time derivatives of the state variables. For example, the vector term in this version features the acceleration, similar to the one proposed by Eckart in a phenomenological fashion~\cite{Eckart}. Up to now, and to the authors' knowledge, it is not possible to obtain such a structure by strictly following the traditional CE method.
Therefore, in the relativistic regime, one obtains two different sets of expressions for the dissipative fluxes. As will be discussed, these two representations are equivalent to each other and correspond to particular cases of the general first-order theories put forward in Refs.~\cite{BDNK_PRX,pK19}.

The remainder of this article is organized as follows. In Section~\ref{Sec:NR} the non-relativistic Boltzmann equation is treated by the two methods described above, leading to the same force-flux relations. Section \ref{Sec:R} is devoted to the relativistic counterpart where it is shown that one obtains two different source terms. At the beginning of both sections we also present the particular form of Eq.~(\ref{eq1-1}) for
non-relativistic and relativistic gases, respectively, together with a dimensional analysis that justifies the CE expansion. Finally, in Section~\ref{Sec:Discussion} we discuss how each representation in the relativistic scenario leads to different constitutive relations, which are nevertheless shown to be equivalent. A thorough discussion of the results and the main conclusions of this article are also included in Section~\ref{Sec:Discussion}. In the appendix we summarize the frame-invariant formulation put forward by Kovtun~\cite{pK19} and extend it to include the electromagnetic force. Furthermore, we provide definitions for the transport coefficients which are representation-independent in addition to frame-invariant within the context of the first-order theories.


\section{Non-relativistic case}

\label{Sec:NR} 

As mentioned above, in this section we address the non-relativistic
version of the Boltzmann equation. In particular, we consider a simple,
mild temperature gas of classical, massive and charged particles in
the presence of an external electromagnetic field. The regime under
consideration, regarding the temperature of the system,
is characterized by large values of the parameter $z=mc^{2}/\kb T$,
which measures the ratio between the rest energy of the individual particles
of mass $m$ ($c$ being the speed of light) and the thermal energy
of the gas $\kb T$ ($\kb$ denoting Boltzmann's constant and $T$
the temperature). Under these assumptions a Newtonian treatment of
the problem is adequate, and the Liouville operator in Eq. (\ref{eq1-1})
is given by 
\begin{equation}
L_{F}\left[f\right]=\frac{\partial f}{\partial t}+\vec{v}\cdot\frac{\partial f}{\partial\vec{r}}+\frac{\vec{F}}{m}\cdot\frac{\partial f}{\partial\vec{v}},\label{eq:LOperator}
\end{equation}
where $\vec{v}$ is the molecular velocity and $\vec{F}$ is the external (electromagnetic) force acting on
the system. The collision term in this case is given by the usual
integral operator 
\begin{equation}
Q\left(ff'\right)=\int\int\left[f\,'f_{1}'-ff_{1}\right]g\sigma\left(g,\Omega\right)d\Omega d^{3}v_{1},\label{eq:Kernel}
\end{equation}
measuring the change in the one-particle distribution function
arising from collisions with molecules labeled with a subscript $1$.
Here primes denote quantities after the collision and the shorthand
notation $f\,'=f\left(\vec{r},\vec{v}\,',t\right)$ has been introduced. Also,
$g$ stands for the norm of the relative velocity, $\sigma$ is the
differential cross section, and $\Omega$ the solid angle. For more
details on the setup of this equation, the reader is referred to Ref.~\cite{Struchtrup}, for example.

In order to compare the order of magnitude of each term in Boltzmann's
equation (\ref{eq1-1}) one
introduces dimensionless variables, denoted by overbars, as follows: $t=\left(L/\vt \right)\bar{t}$,
$\vec{r}=L\vec{\bar{r}},\,$$\vec{v}=\vt \vec{\bar{v}}$,
$g=\sqrt{2}\vt \bar{g}$, $f=\bar{f}/\left(\vt^{3}L^{3}\right)$,
$\sigma=\pi d^{2}\bar{\sigma}$ and $\vec{F}=\left(m\vt^{2}/L\right)\vec{\bar{F}}$. Here $L$ is a characteristic macroscopic length scale, $\vt$ a typical mesoscopic speed and $d$ a microscopic scale of the order of magnitude of the
molecules' diameter. 
Defining the mean free path as
$\ell:=L^{3}/\left(\sqrt{2}\pi d^{2}\right)$ and
introducing these scaled variables in Eq.~(\ref{eq1-1}) one obtains
(omitting the bars) 
\begin{equation}
\frac{\partial f}{\partial t}+\vec{v}\cdot\frac{\partial f}{\partial\vec{r}}+\vec{F}\cdot\frac{\partial f}{\partial\vec{v}}=\frac{1}{\text{Kn}}Q\left(ff'\right),
\label{eq:BoltzmanNR}
\end{equation}
where $\text{Kn}=\ell/L$ is the Knudsen parameter.

Motivated by Eq.~(\ref{eq:BoltzmanNR}), the CE method relies on 
the hypothesis that the distribution
function can be expanded as 
\begin{equation}
f=\sum_{i=0}^{\infty}\text{Kn}^{i}f^{\left(i\right)},\label{eq:expansion}
\end{equation}
where $f^{\left(0\right)}$ is a local (Maxwell-Boltzmann) equilibrium distribution function depending on the state variables, which we choose as the particle number density $n$, the temperature $T$ and the bulk velocity $\vec{u}$. These variables are determined by the usual compatibility conditions involving the first few moments of $f$, and hence they depend themselves on $\text{Kn}$~\cite{Cercignani2012boltzmann,LaurePaper}.

\subsection{The traditional Chapman-Enskog procedure}

\label{CENR}

In the traditional approach the expansion given by Eq.~(\ref{eq:expansion})
is assumed to have a different impact on the convective time derivative
operator than on the rest of the terms on the left-hand side of Eq.~(\ref{eq:BoltzmanNR}). Indeed, separating the molecular velocity $\vec{v}$ in
its chaotic $\vec{k}$ and bulk $\vec{u}$ components as $\vec{v}=\vec{u}+\vec{k}$ one
can rewrite Eq. (\ref{eq:BoltzmanNR}) as
\begin{equation}
\mathcal{D}f+\vec{k}\cdot\frac{\partial f}{\partial\vec{r}}+\frac{\vec{F}}{m}\cdot\frac{\partial f}{\partial\vec{v}}=\frac{1}{\text{Kn}}Q\left(ff'\right),\label{eq:BoltzmannNR2}
\end{equation}
where $\mathcal{D}:=\partial/\partial t+\vec{u}\cdot\left(\partial/\partial\vec{r}\right)$
is the convective derivative. The procedure assumes that
while the expansions for the terms $\left(\partial f/\partial\vec{r}\right)$
and $\left(\partial f/\partial\vec{v}\right)$ read 
\begin{equation}
\frac{\partial f}{\partial\vec{r}}=\sum_{i=0}^{\infty}\text{Kn}^{i}\frac{\partial f^{\left(i\right)}}{\partial\vec{r}},\label{eq:exp1}
\end{equation}
and 
\begin{equation}
\frac{\partial f}{\partial\vec{v}}=\sum_{i=0}^{\infty}\text{Kn}^{i}\frac{\partial f^{\left(i\right)}}{\partial\vec{v}},\label{eq:exp2}
\end{equation}
the first term requires a different expansion, i.e. 
\begin{equation}
\mathcal{D}f=\sum_{i=0}^{\infty}\text{Kn}^{i}\left(\mathcal{D}f\right)_{i},\label{eq:exp3}
\end{equation}
where the coefficients $\left(\mathcal{D}f\right)_{i}$ are still to be determined. This ansatz can be motivated by various arguments. The simplest one consists in noticing that the operator $\mathcal{D}$ involves only macroscopic variables, unlike the other two terms which depend on the molecular velocity. For a thorough discussion of this point the reader is referred to Refs.~\cite{Ch-E} and~\cite{cercigniani1} for example.

Introducing Eqs.~(\ref{eq:exp1})-(\ref{eq:exp3}) into Eq.~(\ref{eq:BoltzmannNR2})
one obtains, for the first two orders in $\text{Kn}$
\begin{equation}
Q\left(f^{\left(0\right)}f^{\left(0\right)'}\right)=0,\label{eq:ordencero}
\end{equation}
and 
\begin{equation}
\left(\mathcal{D}f\right)_{0}+\vec{k}\cdot\frac{\partial f^{\left(0\right)}}{\partial\vec{r}}+\frac{\vec{F}}{m}\cdot\frac{\partial f^{\left(0\right)}}{\partial\vec{v}}=f^{\left(0\right)}C\left(\phi \right),\label{eq:ordenuno}
\end{equation}
where we recall that $f^{\left(1\right)} = f^{\left(0\right)}\phi$ and
\begin{equation}
C\left(\phi\right):=\int f_{1}^{\left(0\right)}\left[\phi'_{1}+\phi'-\phi_{1}-\phi\right]g\sigma\left(g,\Omega\right)d\Omega d^{3}v_{1},\label{C}
\end{equation}
is the linearized collision operator \cite{Kremer-BookNR}. Equation~(\ref{eq:ordencero}) implies that the first term in the expansion is precisely the Maxwell-Boltzmann distribution function 
\begin{equation}
f^{\left(0\right)}=n\left(\frac{m}{2\pi \kb T}\right)^{3/2}\exp\left(-\frac{m\left(\vec{v}-\vec{u}\right)^{2}}{2\kb T}\right).
\label{eq:distMB}
\end{equation}
For the second equation
(Eq. (\ref{eq:ordenuno})) one introduces the local equilibrium, or
functional, hypothesis by means of which the space and time dependence
of the distribution function is given solely through the state variables
that parametrize it. This assumption is also considered for the undetermined
coefficients $\left(\mathcal{D}f\right)_{i}$. In particular for $i=0$,
\begin{equation}
\left(\mathcal{D}f\right)_{0}=\frac{\partial f^{\left(0\right)}}{\partial n}\left(\mathcal{D}n\right)_{0}+\frac{\partial f^{\left(0\right)}}{\partial T}\left(\mathcal{D}T\right)_{0}+\frac{\partial f^{\left(0\right)}}{\partial\vec{u}}\cdot\left(\mathcal{D}\vec{u}\right)_{0}.\label{eq:expDeriva}
\end{equation}
Introducing Eq.~(\ref{eq:expDeriva}) into Eq.~(\ref{eq:ordenuno}) one
is led to 
\begin{equation}
\frac{1}{n}\left[\left(\mathcal{D}n\right)_{0}+\vec{k}\cdot\vec{\nabla} n\right] + \frac{1}{T}\left(\frac{mk^{2}}{2\kb T} - \frac{3}{2}\right)\left[\left(\mathcal{D}T\right)_{0}+\vec{k}\cdot\vec{\nabla} T\right]
 + \frac{m\vec{k}}{\kb T}\cdot\left[\left(\mathcal{D}\vec{u}\right)_{0}+(\vec{k}\cdot\vec{\nabla})\vec{u}-\frac{\vec{F}}{m}\right] = C\left(\phi\right).
\label{eq:EqBoltzman2}
\end{equation}
In order to determine the coefficients $\left(\mathcal{D}n\right)_{0}$, $
\left(\mathcal{D}\vec{u}\right)_{0}$ and  $\left(\mathcal{D}T\right)_{0}$,
we multiply both sides of Eq.~(\ref{eq:EqBoltzman2}) by the vector of collision invariants
$\vec{\psi}=\left(1,\vec{v},\frac{1}{2}mv^{2}\right)^{T}$ and $f^{\left(0\right)}$ and integrate over momentum space, obtaining
\begin{align}
\int \vec{\psi}f^{\left(0\right)}\left\{ \frac{1}{n}\left[\left(\mathcal{D}n\right)_{0}+\vec{k}\cdot\vec{\nabla} n\right] + \frac{1}{T}\left(\frac{mk^{2}}{2\kb T}-\frac{3}{2}\right)\left[ \left(\mathcal{D}T\right)_{0}+\vec{k}\cdot\vec{\nabla} T\right] + \frac{m\vec{k}}{\kb T}\cdot\left[ \left(\mathcal{D}\vec{u}\right)_{0}+(\vec{k}\cdot\vec{\nabla})\vec{u}-\frac{\vec{F}}{m}\right] \right\} d^3 v\nonumber \\
=\int\vec{\psi} f^{\left(0\right)} C\left(\phi\right)d^3 v.\label{eq:ParaDes-1}
\end{align}
The right-hand side of Eq.~(\ref{eq:ParaDes-1}) vanishes due to the
symmetry properties of the linearized collision operator and the fact that all components of $\vec{\psi}$ are collision invariants. Consequently, Eq.~(\ref{eq:ParaDes-1}) implies that (see Ref.~\cite{Ch-E} for details)
\begin{equation}
\left(\mathcal{D}n\right)_{0}=-n\vec{\nabla}\cdot\vec{u},\qquad \left(\mathcal{D}\vec u\right)_{0}=-\frac{\vec{\nabla}{\rm {p}}}{nm}+\frac{\vec{F}}{m}, \qquad \left(\mathcal{D}T\right)_{0}=-\frac{2}{3}T\vec{\nabla}\cdot\vec{u},
\label{eq:euler}
\end{equation}
where ${\rm {p}}=n\kb T$ is the hydrostatic pressure. By substituting
the coefficients $\left(\mathcal{D}n\right)_{0},\left(\mathcal{D}\vec{u}\right)_{0},\left(\mathcal{D}T\right)_{0}$
in Eq. (\ref{eq:EqBoltzman2}) one obtains the following integral
equation 
\begin{equation}
C\left(\phi\right)=\left(\frac{mk^{2}}{2\kb T}-\frac{5}{2}\right)\vec{k}\cdot\frac{\vec{\nabla} T}{T}+\frac{m}{\kb T}\left(\vec{k}\vec{k}:\vec{\nabla}\vec{u}-\frac{k^{2}}{3}\vec{\nabla}\cdot\vec{u}\right),
\end{equation}
which can be further simplified by splitting the velocity gradient
as 
\begin{equation}
\vec{\nabla}\vec{u}=\overleftrightarrow{\sigma}+\overleftrightarrow{\omega}+\frac{1}{3}\mathbb{I}\vec{\nabla}\cdot\vec{u},
\end{equation}
where $\overleftrightarrow{\sigma}$ and $\overleftrightarrow{\omega}$ denote, respectively, the symmetric traceless and antisymmetric parts
of $\vec{\nabla}\vec{u}$. Thus, the final form of the integral equation is
\begin{equation}
C\left(\phi\right)=\frac{1}{T}\left(\frac{mk^{2}}{2\kb T}-\frac{5}{2}\right)\vec{k}\cdot\vec{\nabla} T + \frac{m}{\kb T}\vec{k}\vec{k}:\overleftrightarrow{\sigma}.
\label{eq:a}
\end{equation}

Notice that the method described in this subsection is equivalent
to assuming 
\begin{equation}
\frac{\partial f^{\left(0\right)}}{\partial t}=\frac{\partial f^{\left(0\right)}}{\partial n}\frac{\partial n}{\partial t}+\frac{\partial f^{\left(0\right)}}{\partial T}\frac{\partial T}{\partial t}+\frac{\partial f^{\left(0\right)}}{\partial\vec{u}}\cdot\frac{\partial\vec{u}}{\partial t},
\end{equation}
and writing the time derivatives of the state variables $n$, $T$
and $\vec{u}$ in terms of spatial gradients by means of the Euler
equations. Such a procedure can be justified by recognizing the fact
that the time evolution of state variables is only known once the
distribution function is determined. Thus, it is not possible to specify
these quantities to order greater than zero in $\text{Kn}$ at this
point. 
This method has been standarized such that most authors carry out the CE
procedure by simply eliminating the time derivatives of the state
variables in favor of space gradients via the Euler equations.

\subsection{The projection method}
\label{formalNR}

Next, we summarize a more formal approach which follows the lines of Ref. \cite{LaurePaper}. As a starting point, we consider the first-order linear equation~(\ref{2-2}) and apply the Liouville operator defined in Eq.~(\ref{eq:LOperator}) directly on $\ln f^{\left(0\right)}$, with $f^{\left(0\right)}$ given in Eq.~(\ref{eq:distMB}), which yields
\begin{equation}
L_{F}\left[\ln f^{\left(0\right)}\right]=\frac{1}{n}\left(\frac{\partial n}{\partial t}+\vec{v}\cdot\vec{\nabla} n\right)+\frac{1}{T}\left(\frac{mk^{2}}{2\kb T}-\frac{3}{2}\right)\left(\frac{\partial T}{\partial t}+\vec{v}\cdot\vec{\nabla} T\right)+\frac{m\vec{k}}{\kb T}\cdot\left[\frac{\partial\vec{u}}{\partial t}+(\vec{v}\cdot\vec{\nabla})\vec{u}-\frac{\vec{F}}{m}\right].
\label{eq:Louiville1}
\end{equation}
The solvability of Eq.~(\ref{2-2}) requires the source term $L_{F}[\ln f^{\left(0\right)}]$ to lie in the image of the linearized collision operator $C$ defined in Eq.~(\ref{C}). Since the latter is symmetric with respect to the scalar product
\begin{equation}
\langle \phi,\varphi \rangle := \int \phi\, \varphi f^{\left(0\right)} d^3v,
\end{equation}
such that $\langle \phi, C(\varphi) \rangle = \langle C(\phi), \varphi \rangle = \langle L_{F}[\ln f^{\left(0\right)}], \varphi \rangle$ for all $\varphi$, a necessary condition for this is that the source term is orthogonal to the kernel of $C$, which is spanned by the collision invariants $\vec{\psi}=\left(1,m\vec{v},\frac{1}{2}mv^{2}\right)^{T}$. The projection method consists in taking the projection of the source term in Eq.~(\ref{eq:Louiville1}) orthogonal to $\ker(C)$ to enforce this property. In this way, Eq.~(\ref{2-2}) is replaced with
\begin{equation}
C(\phi) = \left( L_{F}\left[\ln f^{\left(0\right)}\right] \right)^\perp,
\label{perp}
\end{equation}
where the superscript $\perp$ refers to the aforementioned orthogonal projection. In order to perform this step, we first rewrite the source term as 
\begin{align}
L_{F}\left[\ln f^{\left(0\right)}\right] & =\frac{m}{\kb T}\left[\frac{1}{2}\left(v^{2}-2\vec{u}\cdot\vec{v}\right)\vec{v}\cdot\frac{\vec{\nabla} T}{T}+\vec{v}\vec{v}:\vec{\nabla}\vec{u}\right]+K,\label{eq:LouOtrogonal}
\end{align}
where terms in $\ker\left(C\right)$ are now separated and included
in $K$, which is given by
\begin{equation}
K := \frac{1}{n}\left(\frac{\partial n}{\partial t}+\vec{v}\cdot\vec{\nabla} n\right)+\frac{1}{T}\left(\frac{mk^{2}}{2\kb T}-\frac{3}{2}\right)\frac{\partial T}{\partial t}+\frac{1}{T}\left(\frac{mu^{2}}{2\kb T}-\frac{3}{2}\right)\vec{v}\cdot\vec{\nabla} T-\frac{m\vec{u}}{\kb T}\cdot\left(\vec{v}\cdot\vec{\nabla}\right)\vec{u}+\frac{m\vec{k}}{\kb T}\cdot\left(\frac{\partial\vec{u}}{\partial t}-\frac{\vec{F}}{m}\right).\label{eq:Kernel-1}
\end{equation}
Notice that all time derivatives are contained in $K$ and hence they lie in $\mathrm{ker}\left(C\right)$. As a consequence, these terms are projected to zero. To project the remaining terms, we make the ansatz
\begin{equation}
\left(\left(v^{2}-2\vec{v}\cdot\vec{u}\right)\vec{v}\right)^{\perp}=\left(v^{2}-2\vec{v}\cdot\vec{u}\right)\vec{v}+\left(a_{1}+a_{2}\vec{v}\cdot\vec{u}+a_{3}v^{2}\right)\vec{u}+b_{1}\vec{v},\label{vv2perp}
\end{equation}
\begin{equation}
\left(\vec{v}\vec{v}\right)^{\perp}=\vec{v}\vec{v}+\left(c_{1}+c_{2}\vec{v}\cdot\vec{u}+c_{3}v^{2}\right)\vec{u}\vec{u}+d_{1}\left(\vec{u}\vec{v}+\vec{v}\vec{u}\right)+\left(e_{1}+e_{2}\vec{v}\cdot\vec{u}+e_{3}v^{2}\right)\mathbb{I},\label{vvperp}
\end{equation}
where the coefficients $a_{i},b_{i},c_{i},d_{i}$ and $e_{i}$ are
obtained by imposing the orthogonality conditions 
$\langle \vec{\psi}, \left(\left(v^{2}-2\vec{v}\cdot\vec{u}\right)\vec{v}\right)^{\perp} \rangle = 0$ and $
\langle \vec{\psi},\left(\vec{v}\vec{v}\right)^{\perp} \rangle = 0$. This yields
\begin{equation}
\left(\left(v^{2}-2\vec{v}\cdot\vec{u}\right)\vec{v}\right)^{\perp}=\left(k^{2}-\frac{5p}{nm}\right)\vec{k},\label{eq:1ort}
\end{equation}
\begin{equation}
\left(\vec{v}\vec{v}\right)^{\perp}=\vec{k}\vec{k}-\frac{\mathbb{I}}{3}k^{2}.\label{eq:2ort}
\end{equation}
Introduction of Eqs.~(\ref{eq:1ort}) and (\ref{eq:2ort}) into Eq.~(\ref{eq:LouOtrogonal}) gives
\begin{equation}
L_{F}\left[\ln f^{\left(0\right)}\right] = \left(  L_{F}\left[\ln f^{\left(0\right)}\right] \right)^\perp + K',
\end{equation}
where
\begin{align}
\left( L_{F}\left[\ln f^{\left(0\right)}\right] \right)^\perp & =\left(\frac{mk^{2}}{2\kb T} -\frac{5}{2}\right)\vec{k}\cdot\frac{\vec{\nabla} T}{T}+\frac{m}{\kb T}\vec{k}\vec{k}:\overleftrightarrow{\sigma}
\label{eq:b}
\end{align}
and 
\begin{align}
K' & =\frac{1}{n}\left(\frac{\partial n}{\partial t}+\vec{u}\cdot\vec{\nabla} n+n\vec{\nabla} \cdot\vec{u}\right)+\frac{1}{T}\left(\frac{mk^{2}}{2\kb T}-\frac{3}{2}\right)\left(\frac{\partial T}{\partial t}+\vec{u}\cdot\vec{\nabla}  T+\frac{2}{3}T\vec{\nabla} \cdot\vec{u}\right)\nonumber \\
 & +\frac{m\vec{k}}{\kb T}\cdot\left[\frac{\partial\vec{u}}{\partial t}+(\vec{u}\cdot\vec{\nabla}) \vec{u}+\frac{\vec{\nabla} {\rm {p}}}{nm}-\frac{\vec{F}}{m}\right],\label{KpNR}
\end{align}
contains all elements of $L_{F}\left[\ln f^{\left(0\right)}\right]$ belonging to $\mathrm{ker}\left(C\right)$. Therefore, replacing the source term with its orthogonal projection is consistent as long as $K'=0$. However, notice
that this is satisfied if the Euler equations are taken into account in Eq.
(\ref{KpNR}), which is a valid assumption to this order. A more thorough
discussion of this statement can be found in the final section of
this work. Furthermore, notice also that Eq.~(\ref{eq:a}) and Eq.~(\ref{perp}) with (\ref{eq:b}) are identical so that both approaches yield exactly the same result.

It is worthwhile to point out that the projection method leaves only spatial
gradients as driving forces. Already
the first step which consists in isolating
terms which belong to $\ker\left(C\right)$ (see Eq. (\ref{eq:LouOtrogonal}))
eliminates the time derivatives and the gradient of the particle number
density, whereas the second step (projecting onto $\ker\left(C\right)^{\perp}$) 
eliminates the divergence of the velocity and modifies the coefficients
of the thermodynamic forces to finally lead to the correct expression.
Thus, this approach does not require making any assumption regarding
the time derivatives of the state variables and leads to the same
structure for the inhomogeneous part of the equation as in the previous procedure. This concludes the proof that in the non-relativistic scenario both
methods lead to the same equation and thus to the same constitutive
relations. 


\section{Relativistic Case}

\label{Sec:R} 
In this section the relativistic scenario is addressed, in which the thermal energy of the system is comparable or larger than the rest
energy of the particles, such that $z\lesssim 1$. Therefore, we consider a gas of massive, non-degenerate charged particles propagating in a flat spacetime $ds^{2}=\eta_{ab}dx^{a}dx^{b} = -(cdt)^2 + (dx^1)^2 + (dx^2)^2 + (dx^3)^2$ in the presence
of an external electromagnetic field, such that
the Liouville operator reads 
\begin{equation}
L_{F}\left[f\right]=p^{a}\frac{\partial f}{\partial x^{a}}+\frac{q}{c}F^{ab}p_{b}\frac{\partial f}{\partial p^{a}},\label{eq:LOperatorR}
\end{equation}
where $F^{ab}$ is the electromagnetic field tensor and the relativistic kernel is given by 
\begin{equation}
Q\left(ff'\right)=\int\int\left[f\,'f_{1}'-ff_{1}\right]\mathcal{F}\sigma(g,\Omega) d\Omega d^{*}p_{1}.\label{eq:KernelR}
\end{equation}
Here $p^{a}$ denotes the molecules' four momenta, satisfying $p^{a}p_{a}=-mc^{2}$,
and $x^{a}$ are the spacetime coordinates. Also $\mathcal{F}$ is
the invariant flux, $\sigma$ the differential cross section, $\Omega$
the solid angle and we use the abbreviation $d^{*}p=cd^{3}p/\left|p_{0}\right|$ for the Lorentz-invariant volume element.
For more details on the setup of Eqs. (\ref{eq:LOperatorR}) and (\ref{eq:KernelR})
the reader is referred to Refs.~\cite{CercignaniKremer-Book} and \cite{deGrootLibro},
for example.

The order of magnitude for the terms in Eq. (\ref{eq1-1}) can be assessed in
a similar way as in the non-relativistic scenario with the following
additional assumptions: $p^{a}=m\vt \bar{p}^{a}$,
$f=\frac{1}{m^{3}\vt^{3}L^{3}}\bar{f}$, $c = \vt\bar{c}$,
$q=e\bar{q}$, $F^{ab}=\frac{m\vt^{2}}{Le}\bar{F}^{ab}$,
$\mathcal{F}=\sqrt{2}m^{2}\vt^{2}\bar{\mathcal{F}}$,
where $e$ is the elementary charge. Proceeding as in Section \ref{Sec:NR},
one is led to the following dimensionless equation 
\begin{equation}
p^{a}\frac{\partial f}{\partial x^{a}}+\frac{q}{c}F^{ab}p_{b}\frac{\partial f}{\partial p^{a}}=\frac{1}{\text{Kn}}Q\left(ff'\right),\label{eq:dimensionlessR}
\end{equation}
where we have once
again omitted the bars over the dimensionless quantities in order to simplify the notation. Thus, in this scenario
the collisional term also scales one order higher in $1/\text{Kn}$ than the transport part, which suggests the same expansion as in the previous
case: $f=\sum_{i=0}^{\infty}\text{Kn}^{i}f^{\left(i\right)}$. As before,
$f^{\left(0\right)}$ corresponds to a local equilibrium distribution
function, parametrized by the dynamical variables $n$, $T$ and $u^{a}$ (where $u^a$ is the hydrodynamic four-velocity),
and is given by the solution to $Q\left(f^{\left(0\right)}f^{\left(0\right)'}\right)=0$
in Eq.~(\ref{eq:KernelR}), which leads to the Maxwell-Jüttner distribution
function~\cite{fJ11a,CercignaniKremer-Book}: 
\begin{equation}
f^{\left(0\right)}=\frac{n}{4\pi m^{2}c\kb T K_{2}\left(z\right)}\exp\left(\frac{u_{a}p^{a}}{\kb T}\right),
\label{eq:juttner}
\end{equation}
where $K_{j}\left(z\right)$ denotes the modified Bessel function of the
second kind and order $j$.

In the relativistic scenario, the transport equations are given through
a conservation equation for the particle four-flux 
\begin{equation}
N^{a}=\int p^{a}fd^{*}p,\label{eq:Na}
\end{equation}
and a balance equation for the energy-momentum-stress tensor 
\begin{equation}
T^{ab}=\int p^{a}p^{b}fd^{*}p,\label{eq:Tab}
\end{equation}
that is,
\begin{equation}
\frac{\partial N^{a}}{\partial x^{a}}=0\quad\text{and}\quad  
\frac{\partial T^{ab}}{\partial x^{a}}=qF^{ba}N_{a}.
\label{Eq:Balance}
\end{equation}
These equations result from multiplying Eq. (\ref{eq1-1}) by the collision invariants
$1$ and $p^{a}$ and integrating over momentum space. In equilibrium one has $N_{(0)}^{a}=\int p^{a}f^{(0)}d^{*}p=nu^{a}$ and $T_{(0)}^{ab}=\int p^{a}p^{b}f^{(0)}d^{*}p=\frac{n\varepsilon}{c^{2}}u^{a}u^{b} + {\rm p} h^{ab}$, where $\varepsilon$ is the internal energy per particle: 
\begin{equation}
\varepsilon=\frac{m^{2}c^{2}}{n}\int x^{2}f^{\left(0\right)}d^{*}p=\left(\mathcal{G}-\frac{1}{z}\right)mc^{2},
\label{epsilon}
\end{equation}
and where we have defined $x:=\left(p^{a}u_{a}\right)/\left(mc^{2}\right)$, $h^{ab}:=\eta^{ab}+\frac{1}{c^{2}}u^{a}u^{b}$ and $\mathcal{G}:=K_{3}\left(z\right)/K_{2}\left(z\right)$.

However, out of equilibrium, there is
no unique way to associate the components of $N^{a}$ and $T^{ab}$ to the state variables and fluxes in the system. In fact, there is a large
family of valid representations as explained in Refs. \cite{pK19,BDNK_PRX}. In the present
work the so-called Eckart frame is considered. This
particular choice is such that 
\begin{equation}
N^{a}=nu^{a},\label{8}
\end{equation}
and
\begin{equation}
T^{ab}=\frac{n\varepsilon}{c^{2}}u^{a}u^{b}+
\left(\p+\Pi\right)h^{ab}+\frac{1}{c^{2}}Q^{a}u^{b}+\frac{1}{c^{2}}Q^{b}u^{a}+\tau^{ab},
\end{equation}
where $\varepsilon$ is given by Eq.~(\ref{epsilon}). The non-equilibrium quantities $Q^{a}$, $\Pi$ and $\tau^{ab}$ are the heat flux, the trace and the traceless part of the Navier viscous tensor, respectively, and $Q^{a}$ and $\tau^{ab}$ are orthogonal to $u^{a}$.
Therefore, this frame satisfies the matching conditions:
\begin{equation}
N^{a}=N_{\left(0\right)}^{a}=\int p^{a}f^{\left(0\right)}d^{*}p,\label{eq:Neq}
\end{equation}
\begin{equation}
u_{a}u_{b}T^{ab}=u_{a}u_{b}T_{\left(0\right)}^{ab}=m^{2}c^{2}\int x^{2}f^{\left(0\right)}d^{*}p,
\end{equation}
which imply
\begin{equation}
\int p^{a}f^{\left(i\right)}d^{*}p=0,\quad \int x^{2}f^{\left(i\right)}d^{*}p=0,\qquad i>0.
\end{equation}
In this
frame the particle flux is aligned with the hydrodynamic
velocity, such that the particle four-flux has no non-equilibrium contribution
and fixes the variables $u^{a}$ and
$n$ in any regime as the ones determined by the local equilibrium
configuration. The remaining state variable that specifies the local equilibrium in
this frame is the internal energy, as it relates in a one-to-one fashion to the temperature. Whereas in the Newtonian regime the internal energy is directly proportional to $T$, in the relativistic case this relation is nonlinear and given by Eq.~(\ref{epsilon}).

As mentioned above, the Eckart frame is only one of the many possible choices that can be adopted in relativistic non-equilibrium linear thermodynamics. A brief discussion of the transformations that relate them is included in the appendix.

\subsection{The relativistic generalization of the traditional  Chapman-Enskog procedure}

\label{CChE}

To carry out a procedure similar to the one described in Section~\ref{CENR}, we separate the convective derivative $\mathcal{D}$ from the remaining terms
in the relativistic Liouville operator. In this case, splitting $p^a$ in chaotic and bulk components through a simple sum as the one considered in the non-relativistic regime
is not possible. However, one can instead separate the time and space
components of the coordinate derivative as follows:
\begin{equation}
\frac{\partial}{\partial x^{a}}=-\frac{u_{a}u^{b}}{c^{2}}\frac{\partial}{\partial x^{b}}+h_{a}^{b}\frac{\partial}{\partial x^{b}},
\end{equation}
such that
\begin{equation}
p^{a}\frac{\partial f}{\partial x^{a}}=-mxu^{a}\frac{\partial f}{\partial x^{a}}+h^{ab}p_{b}\frac{\partial f}{\partial x^{a}}.\label{eq:derivaR}
\end{equation}
Notice that $u^{a}\frac{\partial f}{\partial x^{a}}$ in the first term on the right-hand side of Eq. (\ref{eq:derivaR})
is the equivalent to the convective time derivative
in the non-relativistic case. Thus it is reasonable to denote this term as $\mathcal{D}f$ and to write the relativistic Liouville operator as
\begin{equation}
L_{F}\left[f\right]=-mx\mathcal{D}f+h^{ab}p_{b}\frac{\partial f}{\partial x^{a}}+\frac{q}{c}F^{ab}p_{b}\frac{\partial f}{\partial p^{a}},
\end{equation}
for the CE method. Proceeding as before, one formally expands
$\mathcal{D}f=\sum_{i=0}^{\infty}\text{Kn}^{i}\left(\mathcal{D}f\right)_{i}$
in the first term and 
\begin{equation}
h^{ab}p_{b}\frac{\partial f}{\partial x^{a}}+\frac{q}{c}F^{ab}p_{b}\frac{\partial f}{\partial p^{a}}=\sum_{i=0}^{\infty}\text{Kn}^{i}\left(h^{ab}p_{b}\frac{\partial}{\partial x^{a}}+\frac{q}{c}F^{ab}p_{b}\frac{\partial}{\partial p^{a}}\right)f^{\left(i\right)},
\end{equation}
for the remaining terms in the Liouville operator. Substitution of the proposed expansions
in Eq. (\ref{eq:dimensionlessR}) leads to 
\begin{equation}
Q\left(f^{\left(0\right)}f^{\left(0\right)'}\right)=0,\label{3}
\end{equation}
and 
\begin{equation}
-mx\left(\mathcal{D}f\right)_{0}+h^{ab}p_{b}\frac{\partial f^{\left(0\right)}}{\partial x^{a}}+\frac{q}{c}F^{ab}p_{b}\frac{\partial f^{\left(0\right)}}{\partial p^{a}}=f^{\left(0\right)}C\left(\phi\right).\label{4}
\end{equation}
Here, once again we use the notation $C\left(\phi\right)$ for the linearized
collision operator which has the same properties as its non-relativistic
counterpart (see for example \cite{deGrootLibro}). Equation~(\ref{3})
implies that $f^{\left(0\right)}$ must be the  Maxwell-Jüttner distribution function
given in Eq. (\ref{eq:juttner}) and, invoking the functional
hypothesis, one obtains from Eq.~(\ref{4})
\begin{align}
C\left(\phi\right) & = \frac{1}{n}\left[ h^{ab}p_{b}\frac{\partial n}{\partial x^{a}}-mx\left(\mathcal{D}n\right)_{0}\right] + \frac{zp_{c}}{mc^{2}}\left[ h^{ab}p_{b}\frac{\partial u^{c}}{\partial x^{a}}-mx\left(\mathcal{D}u^{c}\right)_{0}\right]
\nonumber \\
 & +\frac{1}{T}\left(1-z\mathcal{G}-zx\right)\left[ h^{ab}p_{b}\frac{\partial T}{\partial x^{a}}-mx\left(\mathcal{D}T\right)_{0}\right]
+ \frac{qz}{mc^{3}}F^{ab}p_{b}u_{a}.
\label{eq:cn}
\end{align}
In order to find expressions for the coefficients $\left(\mathcal{D}n\right)_{0}$,
$\left(\mathcal{D}T\right)_{0}$ and $\left(\mathcal{D}u^{c}\right)_{0}$
one can once again use the collision invariants. Indeed, multiplying
Eq. (\ref{eq:cn}) by $\vec\psi:=\left(1,p^{a}\right)^{T}$ and
integrating over momentum space one obtains
\begin{equation}
\left(\mathcal{D}n\right)_{0}=-n\theta,\qquad
\left(\mathcal{D}T\right)_{0}=-\frac{{\rm p}\theta}{nc_{v}},\qquad
\left(\mathcal{D}u^{b}\right)_{0} = -\frac{{\rm D}^{b}{\rm p}}{nm\mathcal{G}} + \frac{q}{c}\frac{F^{ba}u_{a}}{m\mathcal{G}},
\end{equation}
where ${\mathrm{D}}^{b}\left(\quad\right):= h^{ab}\partial\left(\quad\right)/\partial x^{a}$
denotes a purely spatial derivative, $\partial u^{a}/\partial x^{a}=\theta$ the expansion,
and $c_{v}$ the specific heat at constant volume per particle. These equations
have the same structure as the relativistic Euler equations:
\begin{equation}
\dot{n}+n\theta=0,\label{contR}
\end{equation}
\begin{equation}
\dot{u}^{b}+\frac{{\rm {D}^{b}{\rm {p}}}}{nm\mathcal{G}}-\frac{1}{m\mathcal{G}}\frac{q}{c}F^{bc}u_{c}=0,\label{euler}
\end{equation}
\begin{equation}
\frac{\dot{T}}{T}+\frac{\kb}{c_{v}}\theta=0.\label{calorR}
\end{equation}
where $\dot{A}:=u^{a}\frac{\partial A}{\partial x^{a}}$ denotes the proper time derivative of a macroscopic quantity $A$, as measured by observers comoving with $u^a$. Thus, as in the non-relativistic case,
this procedure leads to the same result
as if one would directly substitute the Euler equations
in order to write the time derivatives in terms of spatial gradients, and one obtains
\begin{equation}
C\left(\phi\right)=  \left(1+\frac{x}{\mathcal{G}}\right)\left[\frac{{\rm D}^{b}n}{n}+\left(1-z\mathcal{G}\right)\frac{{\rm D}^{b}T}{T}-\frac{z}{mc^{2}}\frac{q}{c}F^{ba}u_{a}\right]p_{b}+ 
 \left[\left(1-z\mathcal{G}-zx\right)\frac{\kb}{c_{v}}+1\right]mx\theta+\frac{zp_{a}p_{b}}{mc^{2}}{\rm D}^{b}u^{a} .
\end{equation}
Considering the following decomposition for the velocity gradient:
\begin{equation}
\frac{\partial u_{b}}{\partial x^{a}}=-\frac{1}{c^{2}}u_{a}\dot{u}_{b}+\sigma_{ab}+\omega_{ab}+\frac{1}{3}\theta h_{ab},\label{eq:decomposition}
\end{equation}
where $\sigma_{ab}$ is the symmetric traceless component and $\omega_{ab}$
the antisymmetric part, one is led to
\begin{align}
C\left(\phi\right) & = \left(1+\frac{x}{\mathcal{G}}\right)\left[\frac{{\rm D}^{b}n}{n}+\left(1-z\mathcal{G}\right)\frac{{\rm D}^{b}T}{T}-\frac{z}{mc^{2}}\frac{q}{c}F^{ba}u_{a}\right]p_{b}+\frac{zp^{a}p^{b}}{mc^{2}}\sigma_{ab}\nonumber \\
 & +\left[zx^{2}\left(\frac{1}{3}-\frac{\kb}{c_{v}}\right)+x\left(\left(1-z\mathcal{G}\right)\frac{\kb}{c_{v}}+1\right)-\frac{z}{3}\right]m\theta.
\label{eq:aR}
\end{align}
By following the present method, all time derivatives are absent in
the integral equation for $\phi$, and thus they will not
be present in the constitutive equations for the dissipative fluxes.

\subsection{The projection method}

\label{FA}

In the relativistic case, the second approach closely follows the methodology
described in Section~\ref{formalNR}. 
Using Eq.~(\ref{eq:juttner}), the left-hand side of Eq.~(\ref{eq1-1}) yields
\begin{align}
L_{F}\left[\ln f^{\left(0\right)}\right] & =p^{a}\frac{1}{n}\frac{\partial n}{\partial x^{a}}+p^{a}\left(1-z\mathcal{G}-\frac{p^{b}u_{b}}{mc^{2}}z\right)\frac{1}{T}\frac{\partial T}{\partial x^{a}}-\frac{zp_{a}}{mc^{2}}\frac{q}{c}F^{ab}u_{b}\nonumber \\
 & +\frac{zp^{a}p^{b}}{mc^{2}}\left(\sigma_{ab}-\frac{1}{c^{2}}u_{a}\dot{u}_{b}\right)+\frac{1}{3}\frac{z\theta}{mc^{2}}\left(\frac{u_{a}u_{b}}{c^{2}}p^{a}p^{b}-m^{2}c^{2}\right),
\label{eq:Louiville2}
\end{align}
where the decomposition given in Eq. (\ref{eq:decomposition}) was
introduced.

Following the same line of thought as in Section~\ref{formalNR}, one first
isolates terms containing elements of $\ker\left(C\right)$ which
in this case is given by $\mathrm{span}\left\{ 1,p^{a}\right\}$. This step leads to the following expression: \begin{equation}
L_{F}\left[\ln f^{\left(0\right)}\right]=\left(-\frac{u_{b}}{T}\frac{\partial T}{\partial x^{a}}+\sigma_{ab}-\frac{1}{c^{2}}u_{a}\dot{u}_{b}+\frac{\theta}{3}\frac{u_{a}u_{b}}{c^{2}}\right)\frac{zp^{a}p^{b}}{mc^{2}}+K_{R},\label{Llogf}
\end{equation}
where terms in $\ker{\left(C\right)}$ are included in $K_{R}$: 
\begin{equation}
K_{R} := p^{a}\frac{1}{n}\frac{\partial n}{\partial x^{a}}+p^{a}\left(1-z\mathcal{G}\right)\frac{1}{T}\frac{\partial T}{\partial x^{a}}-\frac{mz\theta}{3}-\frac{z}{mc^{2}}\frac{q}{c}p_{a}F^{ab}u_{b}.
\end{equation}
Rearranging terms and decomposing $p^{a}=h_{c}^{a}p^{c}-\frac{p^{c}u_{c}}{c^{2}}u^{a}$ in
the first term on the right-hand side of Eq.~(\ref{Llogf}) one obtains 
\begin{align}
L_{F}\left[\ln f^{\left(0\right)}\right] & =\frac{z}{mc^{2}}\left[\left(\frac{\dot{T}}{T}+\frac{\theta}{3}\right)\frac{u^{a}u^{b}}{c^{2}}-\left(\frac{\dot{u}^{b}}{c^{2}}+\frac{{\rm D}^{b}T}{T}\right)u^{a}+\sigma^{ab}\right]p_{a}p_{b}+K_{R}.\label{eq:LF}
\end{align}
Notice that, in contrast to the non-relativistic calculation, not
all time derivatives terms belong to $\ker{\left(C\right)}$. Moreover, the projection
of $p^{a}p^{b}$ onto $\ker{\left(C\right)}^{\perp}$ only modifies the coefficients,
maintaining these time derivatives in the source term. Indeed, considering
\begin{equation}
\left(p^{a}p^{b}\right)^{\perp}=p^{a}p^{b}+\left(A_{1}+A_{2}x\right)u^{a}u^{b}+B_{1}\left(u^{a}p^{b}+u^{b}p^{a}\right)+\left(C_{1}+C_{2}x\right)h^{ab},
\label{eq:proposal}
\end{equation}
with $z$-dependent coefficients $A_i$, $B_i$, $C_i$,
and imposing the orthogonality conditions 
\begin{equation}
\int\left(\begin{array}{c}
1\\
p^{c}
\end{array}\right)\left(p^{a}p^{b}\right)^{\perp}f^{\left(0\right)}d^{*}p=0,
\end{equation}
leads to
\begin{align}
\left(p^{a}p^{b}\right)^{\perp} & =p^{a}p^{b}+\frac{m^{2}u^{a}u^{b}}{3-\frac{c_v}{\kb}}\left\{ \frac{c_{v}}{\kb} + \left[ 5\mathcal{G}\frac{c_{v}}{\kb} - 3\left(\mathcal{G}+\frac{1}{z}\right)\frac{c_{p}}{\kb}\right]x \right\}
\nonumber \\
 & -m\mathcal{G}\left(u^{a}p^{b}+u^{b}p^{a}\right)
 + \frac{m^{2}c^{2}}{3-\frac{c_v}{\kb}}\left[ 1+\left(\mathcal{G}^{2}z-4\mathcal{G}-z\right)x\right] h^{ab},
\label{eq:prj}
\end{align}
where $c_{p}=\kb+c_{v} = \kb(z^2 + 5z\mathcal{G} - z^2\mathcal{G}^2)$ is the specific heat at constant pressure per particle.
Substitution of Eq. (\ref{eq:prj}) into Eq. (\ref{eq:LF}) 
leads to the following form of the source term
\begin{align}
L_{F}\left[\ln f^{\left(0\right)}\right] & =\frac{z}{mc^{2}}\left[\frac{u_{a}u_{b}}{c^{2}}\left(\frac{\dot{T}}{T}+\frac{\theta}{3}\right)-\left(\frac{\dot{u}_{b}}{c^{2}}+\frac{{\rm D}_{b}T}{T}\right)u_{a}+\sigma_{ab}\right]\left(p^{a}p^{b}\right)^{\perp}+K'_{R},
\label{eq:bR}
\end{align}
where $K'_{R}\in\ker{\left(C\right)}$ is given by 
\begin{align}
K'_{R} & =-mx\left(\frac{\dot{n}}{n}+\theta\right)+\frac{m}{3-\frac{c_{v}}{\kb}}\frac{c_{v}}{\kb}\left(\frac{\dot{T}}{T}+\frac{\kb\theta}{c_{v}}\right)\left(4x-\mathcal{G}zx-z\right) +\frac{z\mathcal{G}}{c^{2}}p_{b}\left(\dot{u}^{b}+\frac{{\rm {D}^{b}{\rm p}}}{nm\mathcal{G}}-\frac{1}{m\mathcal{G}}\frac{q}{c}F^{ba}u_{a}\right).\label{eq:Kp}
\end{align}
Thus, this methodology leads to the following integral equation
 for $\phi$
 \begin{align}
C\left(\phi\right) & =\frac{z}{mc^{2}}\left[\frac{u_{a}u_{b}}{c^{2}}\left(\frac{\dot{T}}{T}+\frac{\theta}{3}\right)-\left(\frac{\dot{u}_{b}}{c^{2}}+\frac{{\rm D}_{b}T}{T}\right)u_{a}+\sigma_{ab}\right]\left(p^{a}p^{b}\right)^{\perp}.
\label{eq:bR1}
\end{align}
As before one has $K'_{R}=0$ if the Euler equations~(\ref{contR}--\ref{calorR}) are imposed, which is justified since the expressions in parentheses in Eq.~(\ref{eq:Kp}) are second order
in this scheme (see Section~\ref{Sec:Discussion}). 
Therefore, eliminating  $K'_R$ when taking the orthogonal projection is not only consistent but also satisfies the necessary condition for the existence of the solution of Eq.~(\ref{eq:bR1}), as explained in Section~\ref{formalNR}. 
It is worthwhile commenting at this point that the relativistic
calculation presents a major difference when compared with the non-relativistic counterpart. Both procedures lead to different structures for the source term of the first-order Boltzmann equation: in the traditional approach only spatial gradients and the electromagnetic force term $\frac{q}{c} F^{ba} u_a$ appear, whereas in the projection method the force term is absent. Moreover, in addition to the spatial derivatives, time derivatives of $T$ and $u^a$ survive the projection.
Consequently, these time derivatives will
be present in the solution for $\phi$, and thus they will also appear
in the constitutive equations. From the point of view of classical
irreversible thermodynamics this may be perceived as an inconsistency
since dissipative fluxes are expected to be driven by spatial gradients
of the state variables. However, in the relativistic framework, the constitutive equations resulting from Eq.~(\ref{eq:bR1}) are not only valid but also equivalent to the ones obtained form Eq.~(\ref{eq:aR}) in Section~\ref{CChE}, in the sense of the general first-order theories. This is explained in detail in the next section.


\section{Discussion}

\label{Sec:Discussion} 
In this work we have analyzed two different ways of implementing the CE method, which we referred to as the traditional procedure and the projection method. Both procedures constitute different ways of imposing orthogonality of the source term of the integral equation to the kernel of the linearized collision operator, which is a necessary condition for the existence of the solution. This step is carried out by adjusting $\mathcal{D}$
in Sections~\ref{CENR} and ~\ref{CChE} for the traditional method, and by explicitly projecting the source term onto the space orthogonal to $\ker(C)$ in Sections~\ref{formalNR} and ~\ref{FA}. In the non-relativistic scenario both lead to the same integral equation, namely Eq.~(\ref{eq:a}) and Eq.~(\ref{perp}) together with Eq.~(\ref{eq:b}). Meanwhile, in the relativistic case, the source terms differ from each other and two versions of the integral equation for $\phi$ are obtained: the traditional procedure leads to Eq.~(\ref{eq:aR}), whereas the projection method yields Eq.~(\ref{eq:bR1}). One would expect both equations to be valid and the resulting constitutive equations to be equivalent to each other within the linear scheme. Indeed, notice that Eq.~(\ref{eq:aR}) can be obtained from Eq.~(\ref{eq:bR1}) and vice versa by imposing the Euler equations~(\ref{euler}) and (\ref{calorR}). The fact that one can go from one
result to the other by using the Euler equations gives some
insight into the kind of transformation that will ultimately relate
the flux-force relations in both scenarios.

It is key for the following argument to recognize that, in addition
to the mathematical aspects of the results of the previous sections, working with one or the other version of the CE method might have important physical implications. Indeed, the constitutive equations obtained from the solution of Eq. (\ref{eq:aR}) in the traditional method can be written as
\begin{equation}
Q^{a} = \tilde{\kappa}_{1}\frac{{\rm D}^{a}n}{n}+\tilde{\kappa}_{2}\frac{{\rm D}^{a}T}{T}+\tilde{\kappa}_{4}\frac{q}{ch}F^{ab}u_{b},\label{d1}
    \end{equation}
\begin{equation}
\Pi=\tilde{\pi}_{3}\theta,\quad \tau^{ab}=-2\eta\sigma^{ab},
\label{pi1}
\end{equation}
whereas Eq.~(\ref{eq:bR1}) in the projection method leads to
\begin{equation}
Q^{a}=\bar{\kappa}_{2}\frac{{\rm D}^{a}T}{T}+\bar{\kappa}_{3}\frac{\dot{u}^{a}}{c^{2}},\label{d2}
\end{equation}
\begin{equation}
\Pi=\bar{\pi}_{2}\frac{\dot{T}}{T}+\bar{\pi}_{3}\theta,\quad \tau^{ab}=-2\eta\sigma^{ab}.
\label{pi2}
\end{equation}
Here, the transport coefficients $\tilde{\kappa}_i$, $\bar{\kappa}_i$, $\tilde{\pi}_i$, $\bar{\pi}_i$ and $\eta$ depend on the temperature and can in principle be computed by inverting $C$ for a particular collision model.

Equations~(\ref{d1})-(\ref{pi2}) raise several non-trivial questions: 
\begin{enumerate}
\item[Q1] Are both force-flux representations valid from the phenomenological point of view?
\item[Q2] If so, how do they relate to each other and how can one interpret the various coefficients involved? 
\item[Q3] Which coefficients or combinations thereof are to be identified
as the thermal conductivity and the shear and bulk viscosities? 
\item[Q4] When coupled to the balance equations~(\ref{Eq:Balance}), do Eqs. (\ref{d1}-\ref{pi1}) and/or Eqs. (\ref{d2}-\ref{pi2}) lead to a physically sound theory for relativistic dissipative fluids regarding hyperbolicity, causality and stability? 
\end{enumerate}
The answers to these questions can be argued
by considering the new proposals for first-order theories of relativistic dissipative fluids~\cite{pK19,BDNK_PRX}. In order to address the first three questions, in Appendix~\ref{App:G} we generalize the work
by Kovtun \cite{pK19} to include the electromagnetic field and the additional freedom of adding terms proportional to the Euler equations (which are zero up to first order). This calculation shows that, even though $N^a$ and $T^{ab}$ can adopt the general form given in Eqs.~(\ref{g1}-\ref{g8}), involving $18$ coefficients, only five linear combinations thereof are frame-invariant and independent of the representation, that is, independent of the addition of the Euler equations. In order to describe these results, for definiteness we adopt the Eckart frame, in which case the first-order constitutive equations have the general form
\begin{align}
Q^{a} &=\kappa_{1}\frac{\mathrm{D}^{a}n}{n}+\kappa_{2}\frac{\mathrm{D}^{a}T}{T}+\kappa_{3}\frac{\dot{u}^{a}}{c^2}+\kappa_{4}\frac{q}{hc}F^{ab}u_{b},
\label{Eq:g7}\\
\Pi &= \pi_{1}\frac{\dot{n}}{n}+\pi_{2}\frac{\dot{T}}{T}+\pi_{3}\theta,\qquad
\tau^{ab}=-2\eta\sigma^{ab},
\label{Eq:g5}
\end{align}
with coefficients $\pi_i$ and $\kappa_i$. Out of the five frame- and representation-invariant quantities, two of them are related to viscous dissipation and can be readily associated with the shear and bulk viscosities $\eta$ and $\zeta$, respectively, where
\begin{equation}
\zeta=\pi_{1}+\frac{\kb}{c_{v}}\pi_{2}-\pi_{3}.
\end{equation}
The remaining three invariant quantities are related to the coefficients $\kappa_i$ appearing in the heat flux and are
\begin{equation}
\mathcal{K}_{i}=\frac{1}{h}\left( \frac{\kb T}{h}\kappa_{3}-\kappa_{i}\right)\,\text{for }i=1,2,\,\quad\text{and}\quad
\mathcal{K}_{4}=-\frac{1}{h}\left(\kappa_{3} + \kappa_{4}\right),
\label{kas}
\end{equation}
where $h:=\varepsilon + \kb T $ is the enthalpy per particle. Moreover, the quantities $\mathcal{K}_{i}$ are further restricted by imposing compatibility with the existence of global equilibrium states, as explained in Appendix~\ref{App:G}. Namely, they must satisfy the relations 
\begin{equation}
\mathcal{K}_{1}=-\frac{\kb T}{\varepsilon}\mathcal{K}_{2},\quad
\mathcal{K}_{4}=-\frac{h}{\kb T}\mathcal{K}_{1},
\label{eq:k1}
\end{equation}
which reduces the number of invariant and independent transport coefficients to three.

After these remarks, we are ready to provide answers to questions Q1--Q3. To this purpose, we first note that both sets of constitutive equations (Eqs.~(\ref{d1},\ref{pi1})
and (\ref{d2},\ref{pi2})) are particular cases of Eqs.~(\ref{Eq:g7},\ref{Eq:g5}). Indeed, Eqs.~(\ref{d1},\ref{pi1}) correspond to these equations with $\kappa_i = \tilde{\kappa}_i$ and $\pi_i = \tilde{\pi}_i$ where 
\begin{equation}
\tilde{\kappa}_{1}=-\frac{\kb T}{\varepsilon}\tilde{\kappa}_{2},\qquad
\tilde{\kappa}_{4}=-\frac{h}{\kb T}\tilde{\kappa}_{1},
\qquad
\tilde{\kappa}_3 = \tilde{\pi}_1 = \tilde{\pi}_2 = 0.
\end{equation}
Here, the first two relations can be inferred directly from Eq.~(\ref{eq:aR}) by taking into account Eq.~(\ref{epsilon}). In particular, the expressions for the non-equilibrium contributions to $T^{ab}$ in terms of the invariants are
\begin{align}
Q^{a} &=-h\mathcal{K}_{2}\left[\frac{{\mathrm{D}}^{a}T}{T}-\frac{\kb T}{\varepsilon}\left(\frac{{\mathrm{D}}^{a}n}{n}-\frac{q}{c\kb T}F^{ab}u_{b}\right)\right],
\label{ka1}\\
\Pi &=  - \zeta\theta,\quad \tau^{ab}=-2\eta\sigma^{ab}.
\label{pitau}
\end{align}
On the other hand, Eqs.~(\ref{d2},\ref{pi2}) have $\kappa_i = \bar{\kappa}_i$ and $\pi_i = \bar{\pi}_i$ with
\begin{equation}
\bar{\kappa}_1 = \bar{\kappa}_4 = \bar{\pi}_1 = 0,\quad
\bar{\kappa}_{2}=\bar{\kappa}_{3},
\quad
\bar{\pi}_3 = \frac{1}{3}\bar{\pi}_2,
\end{equation}
where the last two relations can be inferred from Eq.~(\ref{eq:bR1}). Consequently, Eqs.~(\ref{d2}) and (\ref{pi2}) can be rewritten as
\begin{align}
Q^{a} &=-\frac{h^{2}}{\varepsilon}\mathcal{K}_{2}\left(\frac{{\mathrm{D}}^{a}T}{T}+\frac{1}{c^{2}}\dot{u}^{a}\right),
\label{ka2}\\
\Pi &= \frac{\zeta}{\frac{\kb}{c_v} - \frac{1}{3}}\left( \frac{\dot{T}}{T} + \frac{\theta}{3} \right),\quad \tau^{ab}=-2\eta\sigma^{ab}\label{pitau2}.
\end{align}

Thus, regarding Q1 and Q2, one concludes that both representations are valid and equivalent in the linear case. Indeed, they are related to one another through the balance equations (see Eqs.~(\ref{e2}) and (\ref{e3})), which justifies the freedom of interchanging the acceleration term in Eq. (\ref{ka2}) for a combination of the pressure gradient and the external electromagnetic force at any time using the Euler equation (Eq.~(\ref{euler})) to obtain Eq.~(\ref{ka1}) and vice versa. Moreover, it is easy to verify that condition~(\ref{eq:k1}) for the existence of global equilibria is satisfied in both cases.

The question Q3 of which coefficients correspond to the thermal conductivity and viscosities
is then straightforward to address. In the non-relativistic case,
one identifies these quantities as the ones involved in the Fourier and Navier-Stokes
laws~\cite{Struchtrup}: 
\begin{equation}
\vec{q}=-\kappa\frac{\vec{\nabla} T}{T},\label{f}
\end{equation}
\begin{equation}
\overleftrightarrow{\pi}=-2\eta\overleftrightarrow{\sigma}-\zeta \mathbb{I}\vec{\nabla}\cdot\vec{u},
\label{pi}
\end{equation}
where $\vec{q}$ and $\overleftrightarrow{\pi}$ are the heat flux and viscous tensor respectively.\footnote{Note that $\zeta=0$ for the Newtonian monoatomic ideal gas considered here.} Whereas Eq.~(\ref{pi}) has precisely the same form as its relativistic counterpart in  Eq.~(\ref{pitau}), the relativistic generalization of the Fourier law is less straightforward. Indeed, if one were to inspect for the possibility of obtaining an expression similar to Eq.~(\ref{f}) in the relativistic scenario one would need to set $\kappa_1 = \kappa_3 = \kappa_4 = 0$ in Eq.~(\ref{Eq:g7}). However, this would lead to a contradiction, since compatibility with global equilibrium would imply that also $\kappa_2 = 0$ (see Eq.~(\ref{eq:k1})). An alternative argument
which allows one to establish the correct expression for the thermal conductivity is based on the
structure of the entropy production, which is given by~\cite{Eckart}
\begin{equation}
\frac{\partial S^a}{\partial x^a} = 
  - \frac{\Pi}{T}\theta - \frac{\tau^{ab}}{T} \sigma_{ab}
  - \frac{Q^a}{T}\left( \frac{{\mathrm{D}}_a T}{T} + \frac{1}{c^2}\dot{u}_a \right),
\end{equation}
where $S^a = n s u^a + \frac{Q^a}{T}$ is the entropy flux and $s = s(\varepsilon,n)$ is the entropy per particle which satisfies the Gibbs relation $Tds = d\varepsilon + {\rm p} d(1/n)$. Requiring positiveness of the term involving the heat flux leads to choosing the constitutive equation~(\ref{ka2}) and the identification of
\begin{equation}
\kappa=\frac{h^{2}}{\varepsilon}\mathcal{K}_{2},
\end{equation}
as the correct thermal conductivity
in order to assure the non-equilibrium version of the second law is exactly satisfied in the absence of other dissipative fluxes. However, notice that the expressions for the force-flux couplings in Eqs.~(\ref{ka1}-\ref{pitau}) and Eqs.~(\ref{ka2}-\ref{pitau2}) also lead to a positive entropy production \textit{when truncated to second order}~\cite{BDNK_PRX}:
\begin{equation}
\frac{\partial S^a}{\partial x^a} = \frac{\Pi^2}{T\zeta} + \frac{\tau^{ab}\tau_{ab}}{2T\eta}
+ \frac{Q^a Q_a}{T\kappa} + \mathcal{E},
\end{equation}
where $\mathcal{E}$ is a correction term which, for the first set (Eqs.~(\ref{ka1}-\ref{pitau})) is
\begin{equation}
\mathcal{E} = -\frac{Q_a}{T}\left( \frac{\dot{u}^a}{c^2} + \frac{\mathrm{D}^a{\rm p}}{n h} 
 - \frac{q}{hc} F^{ab} u_b \right),
\end{equation}
and in the second case (Eqs.~(\ref{ka2}-\ref{pitau2})) reads
\begin{equation}
\mathcal{E} = -\frac{\Pi}{T\left( \frac{\kb}{c_v} - \frac{1}{3}\right)}\left( \frac{\dot{T}}{T} + \frac{\kb}{c_v}\theta \right).
\end{equation}
In both cases we see that, due to balance equations~(\ref{e2},\ref{e3}), $\mathcal{E}$ is third order in the gradients, such that the entropy production is positive up to and including quadratic terms.\footnote{ The correction term $\mathcal{E}$ would be exactly zero for the "mixed" option, Eqs.~(\ref{ka2}) and (\ref{pitau}), corresponding to the Eckart theory; however this cannot be obtained within the CE approach depicted in this article.}

Finally, to address Q4 one needs to resort to the
stability and causality analysis of the system of equations (\ref{Eq:Balance})
when coupled to the linear constitutive relations (\ref{ka1}-\ref{pitau}) or (\ref{ka2}-\ref{pitau2}). Recent works point
to the possibility of obtaining a hyperbolic and causal theory in
a general frame and representation, provided a non-trivial set of inequalities for the  coefficients appearing in the generalized  constitutive equations is satisfied \cite{BDNK_PRX,pK19}. The particular case of the theory obtained in Section~\ref{CChE} has been previously analyzed and shown to lead
to a stable system that does not violate causality in the
comoving frame but becomes unstable in a general one \cite{reularubio}.
A thorough analysis of the theory involving constitutive equations (\ref{ka2}-\ref{pitau2})  is beyond the scope of this article and will be addressed elsewhere. 

Although for the sake of of clarity of the presentation we have restricted ourselves to the special-relativistic scenario in three space dimensions, our results can be easily generalized to a kinetic gas propagating on a curved spacetime of arbitrary dimensions. Notice that the results in Sections~\ref{CChE} and \ref{FA}, regarding the different structure in the integral equations~(\ref{eq:aR}) and Eq.~(\ref{eq:bR1}), are independent of the Eckart frame described at the beginning of that section. Therefore, it would be interesting to analyze whether our findings about the resulting differences in the constitutive equations~(\ref{ka1}-\ref{pitau}) and (\ref{ka2}-\ref{pitau2}), hold in frames more general than Eckart's, based on the study of the homogeneous solution of the first-order equation similar to what has been performed in Ref.~\cite{rHpK22} in the context of the Hilbert expansion.

\acknowledgments 
  
We thank Oscar Reula for fruitful discussions. O.S. was partially supported by CIC Grant No.~18315 to Universidad Michoacana and by CONAHCyT Network Project No. 376127 ``Sombras, lentes y ondas gravitatorias generadas
por objetos compactos astrofísicos".


\appendix


\section{}

\label{App:G} 
In this appendix, the key results used
in the discussion part of the present work are briefly summarized. They are based on Ref.~\cite{pK19}, here extended to include the electromagnetic field contribution and to consider in a general manner the transformations involving the Euler equations.
The key idea behind the general first-order theories formalism relies on the fact that, to first-order in derivatives,
the particle four flux and energy-momentum-stress tensor can be written as
\begin{equation}
J^{a}=Nu^{a}+\mathcal{J}^{a},\label{g1}
\end{equation}
and
\begin{equation}
T^{ab}=\frac{1}{c^{2}}Eu^{a}u^{b}+Ph^{ab}+\frac{1}{c^{2}}Q^{a}u^{b}+\frac{1}{c^{2}}Q^{b}u^{a}+\tau^{ab},\label{g2}
\end{equation}
with 
\begin{equation}
N=n+\nu_{1}\frac{\dot{n}}{n}+\nu_{2}\frac{\dot{T}}{T}+\nu_{3}\theta+\mathcal{O}\left(\partial^{2}\right),\label{g3}
\end{equation}
\begin{equation}
E=n\varepsilon+\varepsilon_{1}\frac{\dot{n}}{n}+\varepsilon_{2}\frac{\dot{T}}{T}+\varepsilon_{3}\theta+\mathcal{O}\left(\partial^{2}\right),\label{g4}
\end{equation}
\begin{equation}
P=\p+\pi_{1}\frac{\dot{n}}{n}+\pi_{2}\frac{\dot{T}}{T}+\pi_{3}\theta+\mathcal{O}\left(\partial^{2}\right),\label{g5}
\end{equation}
\begin{equation}
\mathcal{J}^{a}=\gamma_{1}\frac{\mathrm{D}^{a}n}{n}+\gamma_{2}\frac{\mathrm{D}^{a}T}{T}+\gamma_{3}\frac{\dot{u}^{a}}{c^2}+\gamma_{4}\frac{q}{hc}F^{ab}u_{b}+\mathcal{O}\left(\partial^{2}\right),\label{g6}
\end{equation}
\begin{equation}
Q^{a}=\kappa_{1}\frac{\mathrm{D}^{a}n}{n}+\kappa_{2}\frac{\mathrm{D}^{a}T}{T}+\kappa_{3}\frac{\dot{u}^{a}}{c^2}+\kappa_{4}\frac{q}{hc}F^{ab}u_{b}+\mathcal{O}\left(\partial^{2}\right),\label{g7}
\end{equation}
\begin{equation}
\tau^{ab}=-2\eta\sigma^{ab}+\mathcal{O}\left(\partial^{2}\right),\label{g8}
\end{equation}
where $\mathcal{O}\left(\partial^{2}\right)$ refers to terms being at least quadratic in derivatives of the state variables $n$,
$T$ and $u^{a}$. The $18$ coefficients $\nu_i$, $\varepsilon_i$, $\pi_i$, $\gamma_i$, $\kappa_i$ and $\eta$ introduced are functions of the temperature and quantify the linear corrections to the local equilibrium
configuration. In Ref.~\cite{pK19} it is shown how different choices of frames
lead to different values of the coefficients. However, particular
combinations remain invariant under these transformations.
These quantities are $\eta$,
\begin{equation}
\ell_{i}=\gamma_{i}-\frac{n}{n\varepsilon+\p}\kappa_{i},\quad i=1,2,3,4,
\end{equation}
and
\begin{equation}
f_{i}=\pi_{i}-\frac{1}{nc_{v}}\left(\frac{\partial \p}{\partial T}\right)_{n}\varepsilon_{i}-\left[\left(\frac{\partial \p}{\partial n}\right)_{T}-\frac{1}{nc_{v}}\left(\frac{\partial \p}{\partial T}\right)_{n}\left(\frac{\partial n\varepsilon}{\partial n}\right)_{T}\right]\nu_{i},\quad i=1,2,3;
\end{equation}
hence only $8$ combinations out of the $18$ coefficients are frame-independent.

In the particular case of this work, where the Eckart frame is chosen
and an ideal gas with $\p=nkT$ is considered, one has $\gamma_{i}=\nu_i=\epsilon_i=0$ 
and thus, the invariant combinations in this frame are 
\begin{equation}
f_{i}:=\pi_{i},\, i=1,2,3,\quad\ell_{i}:=-\frac{1}{h}\kappa_{i},\, i=1,2,3,4,
\end{equation}
and the particle flux and energy-momentum-stress tensor are given by $J^{a}=nu^{a}$ and
\begin{equation}
 T^{ab}=\frac{1}{c^{2}}n\varepsilon u^{a}u^{b}+Ph^{ab}+\frac{1}{c^{2}}Q^{a}u^{b}+\frac{1}{c^{2}}Q^{b}u^{a}+\tau^{ab},
\end{equation}
with $P$, $Q^{a}$ and $\tau^{ab}$ given by Eqs. (\ref{g5}), (\ref{g7})
and (\ref{g8}). There is an additional freedom in the linear scenario which leads to different representations within the same frame.
Indeed, the balance equations imply that 
\begin{equation}
\frac{\dot{n}}{n}+\theta=\mathcal{O}\left(\partial^{2}\right),\label{e1}
\end{equation}
\begin{equation}
\frac{1}{c^{2}}\dot{u}_{a}+\frac{\kb T}{h}\left(\frac{\mathrm{D}_an}{n}+\frac{\mathrm{D}_aT}{T}\right)-\frac{1}{h}\frac{q}{c}F_{a}{}^{b}u_{b}=\mathcal{O}\left(\partial^{2}\right).\label{e2}
\end{equation}
\begin{equation}
\frac{\dot{T}}{T}+\frac{\kb}{c_{v}}\theta=\mathcal{O}\left(\partial^{2}\right),\label{e3}
\end{equation}
Thus, multiplying Eq.
(\ref{e1}) by $\Omega$, Eq. (\ref{e2}) by $\chi$ and Eq. (\ref{e3})
by $\xi$, and adding them to Eqs.~(\ref{g5}) and (\ref{g7}), one
sees that the coefficients $\pi_i$ and $\kappa_i$ transform according to 
\begin{equation}
\hat{\pi}_{1}=\pi_{1}+\Omega,\quad
\hat{\pi}_{2}=\pi_{2}+\chi,\quad
\hat{\pi}_{3}=\pi_{3}+\Omega+\frac{\kb}{c_{v}}\chi,
\end{equation}
and
\begin{equation}
\hat{\kappa}_{1}=\kappa_{1}+\frac{\kb T}{h}\xi,\quad
\hat{\kappa}_{2}=\kappa_{2}+\frac{\kb T}{h}\xi,\quad
\hat{\kappa}_{3}=\kappa_{3}+\xi,\quad
\hat{\kappa}_{4}=\kappa_{4}-\xi.
\end{equation}
This transformation further reduces the number of invariant coefficients from $8$ to $5$, namely
\begin{equation}
\eta,\quad
\zeta=\pi_{1}+\frac{\kb}{c_{v}}\pi_{2}-\pi_{3},\quad
\mathcal{K}_{i}=\ell_{i}-\frac{\kb T}{h}\ell_{3}\,\quad\text{for }i=1,2\,\quad\text{and}\quad
\mathcal{K}_{4}=\ell_{4} + \ell_{3},
\end{equation}
which are independent of both the frame and the choice of $\Omega$, $\chi$ and $\xi$ (which we refer to as a choice of representation in this article). The number of independent coefficients is further reduced by recognizing
that all descriptions
are required to coincide with each other in global equilibrium. 
Noticing that this condition demands collisions not
to affect the occupation number in phase space, leads to 
\begin{equation}
    p^{\mu}\frac{\partial f^{\left(0\right)}}{\partial x^{\mu}}+\frac{q}{c}F^{ab}p_{b}\frac{\partial f^{\left(0\right)}}{\partial p^{a}}=0,
\end{equation}
which implies
\begin{equation}
-\frac{\mathrm{D}^{a}T}{T} = \frac{\dot{u}^{a}}{c^2},\quad
\frac{\mathrm{D}_{a}n}{n}=\frac{\varepsilon}{\kb T}\frac{\mathrm{D}_{a}T}{T}+\frac{1}{\kb T}\frac{q}{c}F_{ab}u^{b}.\label{s}
\end{equation}
These restrictions, together with the requirement that the heat flux vanishes in equilibrium, yield the following compatibility
conditions in this framework:
\begin{equation}
\frac{\varepsilon}{\kb T}\kappa_{1} + \kappa_{2} - \kappa_{3}=0,\quad
\kappa_{4} + \frac{h}{\kb T}\kappa_{1}=0.
\label{compkappa}
\end{equation}
By taking into account Eq. (\ref{compkappa}), one is finally left with only three frame- and representation-independent quantities, namely the viscous coefficients $\eta$, $\zeta$
and either $\mathcal{\mathcal{K}}_{1}$, $\mathcal{K}_{2}$ or $\mathcal{K}_{4}$
since they are related by
\begin{equation}
\mathcal{K}_{1}=-\frac{\kb T}{\varepsilon}\mathcal{K}_{2},\quad
\mathcal{K}_{4}=-\frac{h}{\kb T}\mathcal{K}_{1}.
\end{equation}
These results provide the basis for the discussion section in this
work.


\bibliographystyle{ieeetr.bst}
\bibliography{refs_kinetic}
\end{document}